\documentclass{PoS}

\usepackage{units}
\usepackage{amsmath}
\usepackage{amssymb}
\usepackage{graphicx}

\renewcommand\[{\begin{equation}}
\renewcommand\]{\end{equation}}
\numberwithin{equation}{section}

\makeatletter
\let\origsection\section
\renewcommand\section{\@ifstar{\starsection}{\nostarsection}}

\newcommand\nostarsection[1]
{\sectionprelude\origsection{#1}\sectionpostlude}

\newcommand\starsection[1]
{\sectionprelude\origsection*{#1}\sectionpostlude}

\newcommand\sectionprelude{%
  \vspace{-0.4em}
}

\newcommand\sectionpostlude{%
  \vspace{-0.4em}
}
\makeatother

\makeatletter

\providecommand{\tabularnewline}{\\}
\newcommand{\lyxdot}{.}

\makeatother

\title{Long distance contributions to the rare kaon decay $K\rightarrow\pi\ell^{+}\ell^{-}$}

\ShortTitle{$K\rightarrow\pi\ell^{+}\ell^{-}$}

\author{Norman Christ$^a$, Xu Feng$^a$, Andreas J\"{u}ttner$^b$, \speaker{Andrew Lawson}$^b$, Antonin Portelli$^b$, Christopher Sachrajda$^b$\\
        \llap{$^a$}Physics Department, Columbia University, New York NY 10027, USA\\
        \llap{$^b$}School of Physics and Astronomy, University of Southampton Southampton, SO17 1BJ, UK\\
        Email: \email{nhc@phys.columbia.edu}, \email{pkufengxu@gmail.com}, \email{juettner@soton.ac.uk}, \email{al1g13@soton.ac.uk}, \email{antonin.portelli@me.com}, \email{cts@soton.ac.uk}}

\abstract{The rare decays of a kaon into a pion and a charged lepton/antilepton pair proceed via a flavour changing neutral current and therefore may only be induced beyond tree level in the Standard Model. This natural suppression makes these decays sensitive to the effects of potential New Physics. To discern such New Physics one must be able to control the errors on the Standard Model prediction of the decay amplitude. These particular decay channels however are dominated by a single photon exchange; this involves a sizeable long-distance hadronic contribution which represents the current major source of theoretical uncertainty. Here we outline our methodology for the computation of the long distance contributions to these rare decay amplitudes using lattice QCD, and present the numerical results of some exploratory studies using the Domain Wall Fermion ensembles of the RBC and UKQCD collaborations.}

\FullConference{The 33rd International Symposium on Lattice Field Theory\\
		14 -18 July 2015\\
		Kobe International Conference Center, Kobe, Japan*}

\begin{document}

\section{Introduction}

The processes $K\rightarrow\pi\ell\bar{\ell}$ are interesting phenomenologically
as they involve a flavour changing neutral current (FCNC). FCNCs are
forbidden at tree level in the Standard Model (SM); the natural suppression
of $K\rightarrow\pi\ell\bar{\ell}$ amplitudes makes them sensitive
to potential New Physics. These decays may also be used to determine
SM parameters such as $V_{td}$ and $V_{ts}$, study CP violation
and to test chiral perturbation theory (ChPT) descriptions of QCD
at low energies~\cite{Cirigliano:2011ny}.

The decays $K^{+}\rightarrow\pi^{+}\ell^{+}\ell^{-}$ and $K_{S}\rightarrow\pi^{0}\ell^{+}\ell^{-}$
are dominated by long distance effects induced by the one-photon exchange
amplitude \cite{Isidori:2005tv,Christ:2015aha}. There is a significant contribution to this amplitude in the region
where the photon emission and the $W$ exchange are separated by distances
as large as $1/\Lambda_{QCD}$. Such long distance effects contain
significant non-perturbative contributions, and so naturally we can
use lattice QCD to evaluate them.

We aim to compute the Minkowski amplitude of the process $K\rightarrow\pi\gamma^{*}$,
i.e. 
\begin{equation}
\mathcal{A}_{\mu}\left(q^{2}\right)=\int d^{4}x\left\langle \pi\left(\mathbf{p}\right)|T\left[J_{\mu}\left(0\right)H_{W}\left(x\right)\right]|K\left(\mathbf{k}\right)\right\rangle .\label{eq:Minkowski_amplitude}
\end{equation}
Using electromagnetic gauge invariance this non-local
matrix element can be written as
\[
\mathcal{A}_{\mu}\left(q^{2}\right)\equiv\dfrac{V\left(z\right)}{4\pi^{2}}\left(q^{2}\left(k+p\right)_{\mu}-\left(m_{K}^{2}-m_{\pi}^{2}\right)q_{\mu}\right),
\]
where non-perturbative QCD effects are contained in the form factor
$V\left(z\right)$, $z=q^{2}/m_{K}^{2}$.
This form factor has been parametrised in the context of chiral perturbation theory, in the
form~\cite{Ecker:1987qi, D'Ambrosio:1998yj}
\begin{equation}
V(z)=a+bz+V_{\pi\pi}\left(z\right),
\end{equation}
The component $V_{\pi\pi}\left(z\right)$ represents the contribution
coming from $\gamma^{*}\rightarrow\pi\pi$ effects. The contribution
of excited states is encapsulated in the polynomial term $a+bz$, the
coefficients of which have previously been determined from experimental data~\cite{Cirigliano:2011ny}. One
opportunity of lattice QCD is to test this relation by determining
the constants $a$ and $b$ from simulation data.

This paper is organised as follows. In section \ref{sec:Lattice-Methodology}
we discuss the implementation of this decay on the lattice. In section
\ref{sec:Analysis} we go on to discuss the analysis of the correlators
computed on the lattice such that the matrix element of the decay
may be extracted. We discuss the numerical results of our exploratory
studies in section \ref{sec:Numerical-Results}. Finally in section
\ref{sec:Conclusions} we make our conclusions.

\section{Lattice Methodology\label{sec:Lattice-Methodology}}

In order to extract the decay amplitude on the lattice we measure the `unintegrated' 4pt correlator~\cite{Christ:2015aha}
\[
\Gamma_{\mu}^{\left(4\right)}\left(t_{H},t_{J},\mathbf{k},\mathbf{p}\right)=\sum_{\mathbf{x},\mathbf{y}}e^{-i\mathbf{q}\cdot\mathbf{x}}\left\langle \phi_{\pi}\left(t_{\pi},\mathbf{p}\right)T\left[J_{\mu}\left(t_{J},\mathbf{x}\right)H_{W}\left(t_{H},\mathbf{y}\right)\right]\phi_{K}^{\dagger}\left(0,\mathbf{k}\right)\right\rangle ,
\]
where the operator $\phi_{P}\left(t,\mathbf{p}\right)$ is the annihilation operator for a pseudoscalar meson $P$ with momentum $\mathbf{p}$ at a time $t$.
The effective weak Hamiltonian relevant to the transition $s\rightarrow d\ell^{+}\ell^{-}$,
renormalised at a scale $M_{W}\gg\mu>m_{c}$, is defined by~\cite{Buchalla:1995vs}
\begin{equation}
H_{W}=\dfrac{G_{F}}{\sqrt{2}}V_{us}^{*}V_{ud}\left(\sum_{j=1}^{2}C_{j}\left(Q_{j}^{u}-Q_{j}^{c}\right)+\sum_{j=3}^{8}C_{j}Q_{j}+\mathcal{O}\left(\dfrac{V_{ts}^{*}V_{td}}{V_{us}^{*}V_{ud}}\right)\right).\label{eq:weak_hamiltonian}
\end{equation}
In our implementation we consider only the operators $Q_{1}^{u,c}$
and $Q_{2}^{u,c}$;  all others vanish at tree level and have much smaller Wilson coefficients than these two. The operators $Q_{1,2}^{q}$ are defined as~\cite{Isidori:2005tv,Christ:2015aha}
\begin{eqnarray*}
Q_{1}^{q}=\left(\bar{s}_{i}\gamma_{\mu}^{L}d_{i}\right)\left(\bar{q}_{j}\gamma^{L,\mu}q_{j}\right) & \& & Q_{2}^{q}=\left(\bar{s}_{i}\gamma_{\mu}^{L}d_{j}\right)\left(\bar{q}_{j}\gamma^{L,\mu}q_{i}\right),
\end{eqnarray*}
where $i,j$ are summed colour indices and $\gamma_{\mu}^{L}=\gamma_{\mu}\left(1-\gamma_{5}\right)$. To make contact with the continuum we first must non-perturbatively renormalise our lattice operators; we can then use perturbation theory to match with the Wilson coefficients for the $\overline{\mathrm{MS}}$ scheme, which are known at NLO~\cite{Buchalla:1995vs}.
The electromagnetic current $J_{\mu}$ is taken to be the standard
flavour-diagonal current
\[
J_{\mu}=\dfrac{1}{3}\left(2V_{\mu}^{u}-V_{\mu}^{d}-V_{\mu}^{s}+2V_{\mu}^{c}\right),
\]
where $V_{\mu}^{q}$ is the conserved vector current for the flavour $q$. As this current is conserved it requires no renormalisation.

\begin{center}
\begin{figure}
\begin{centering}
\begin{tabular}{cccc}
\includegraphics[scale=0.5]{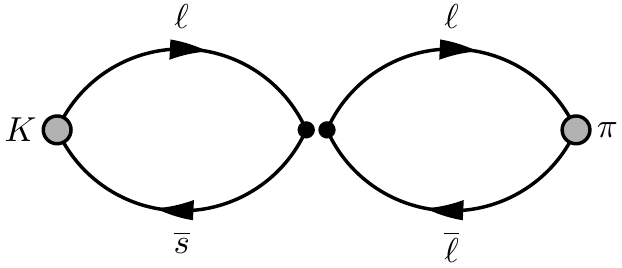} & \includegraphics[scale=0.5]{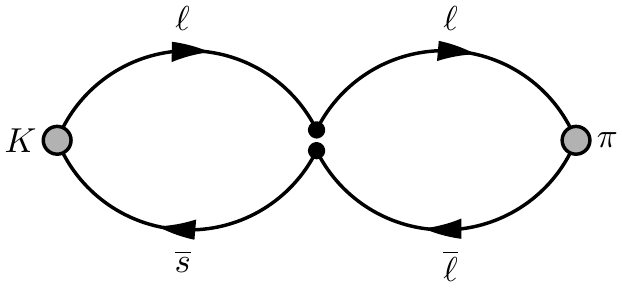} & \includegraphics[scale=0.5]{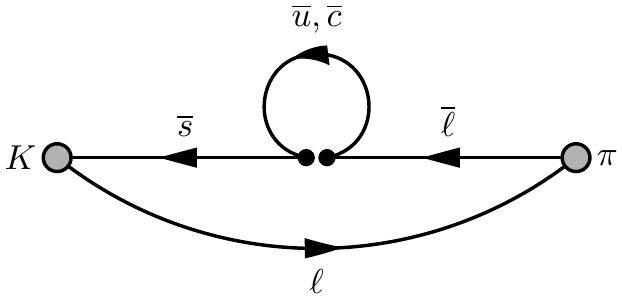} & \includegraphics[scale=0.5]{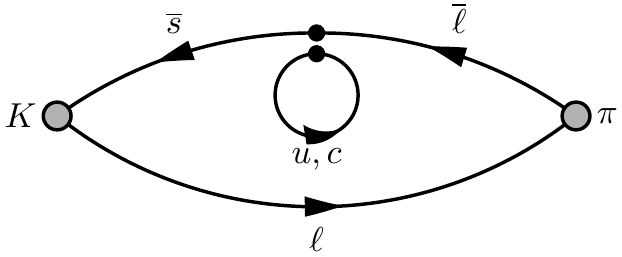}\tabularnewline
$W$ & $C$ & $S$ & $E$\tabularnewline
\end{tabular}
\par\end{centering}

\protect\caption{\label{fig:H_W_contractions}The four diagram topologies obtained
after performing the Wick contractions for the $H_{W}$ operator.}
\end{figure}
\vspace{-2em}

\par\end{center}

In Fig. \ref{fig:H_W_contractions} we display the diagram topologies obtained by performing Wick contractions for just the $H_{W}$ operator. 
The current can be inserted on any of the quark propagators in
each class; however there is also the possibility of the self-contraction
of the current to produce a disconnected diagram, corresponding to a sea quark loop emitting a photon. The full list of diagrams can be found in Ref.~\cite{Christ:2015aha}. When the current is inserted in the loop of the $S$ and $E$ diagrams, they appear quadratically divergent as the operators $J_{\mu}$ and $H_{W}$ approach each other. However, with gauge invariance and the GIM mechanism it can be shown explicitly~\cite{Isidori:2005tv,Christ:2015aha} that these diagrams introduce no new divergences.

\section{Analysis\label{sec:Analysis}}

To recover the amplitude of the decay we consider the integrated 4pt
correlator,
\[
I_{\mu}\left(T_{a},T_{b},\mathbf{k},\mathbf{p}\right)=e^{-\left(E_{\pi}\left(\mathbf{p}\right)-E_{K}\left(\mathbf{k}\right)\right)t_{J}}\int_{t_{J}-T_{a}}^{t_{J}+T_{b}}dt_{H}\tilde{\Gamma}_{\mu}^{\left(4\right)}\left(t_{H},t_{J},\mathbf{k},\mathbf{p}\right)
\]
in the limit $T_{A},T_{B}\rightarrow\infty$, where
\[
\tilde{\Gamma}_{\mu}^{\left(4\right)}=\dfrac{\Gamma_{\mu}^{\left(4\right)}}{Z_{\pi K}},\,Z_{\pi K}=\dfrac{Z_{\pi}Z_{K}^{\dagger}}{4E_{\pi}\left(\mathbf{p}\right)E_{K}\left(\mathbf{k}\right)}e^{-t_{\pi}E_{\pi}\left(\mathbf{p}\right)}.
\]

The spectral decomposition for the integrated 4pt correlator is
\begin{eqnarray}
I_{\mu}\left(T_{A},T_{B},\mathbf{k},\mathbf{p}\right) & = & -\sum_{n}\dfrac{1}{2E_{n}}\dfrac{\left\langle \pi\left(\mathbf{p}\right)|J_{\mu}|n,\mathbf{k}\right\rangle \left\langle n,\mathbf{k}|H_{W}|K\left(\mathbf{k}\right)\right\rangle }{E_{K}\left(\mathbf{k}\right)-E_{n}}\left(1-e^{\left(E_{K}\left(\mathbf{k}\right)-E_{n}\right)T_{A}}\right)\nonumber \\
 &  & +\sum_{m}\dfrac{1}{2E_{m}}\dfrac{\left\langle \pi\left(\mathbf{p}\right)|H_{W}|m,\mathbf{p}\right\rangle \left\langle m,\mathbf{p}|J_{\mu}|K\left(\mathbf{k}\right)\right\rangle }{E_{m}-E_{\pi}\left(\mathbf{p}\right)}\left(1-e^{-\left(E_{m}-E_{\pi}\left(\mathbf{p}\right)\right)T_{B}}\right)\label{eq:int_spec_rep-1}
\end{eqnarray}
The states $m$ must have the same quantum numbers of a kaon, and
thus all possible states will have $E_{m}>E_{\pi}\left(\mathbf{p}\right)$;
this half of the integral thus converges as $T_B\rightarrow\infty$.
However the states $n$ have the
quantum numbers of a pion. For physical pion and kaon masses there
are three permitted on-shell intermediate states (namely one, two
and three pion intermediate states), which will cause the integral
to diverge with increasing $T_{A}$. These on-shell intermediate states
do not contribute to the overall decay width and therefore must be
removed in order to extract the relevant Minkowski amplitude,
\[
\mathcal{A}_{\mu}\left(q^{2}\right)=-i\lim_{T_{a},T_{b}\rightarrow\infty}\tilde{I}_{\mu}\left(T_{A},T_{B},\mathbf{k},\mathbf{p}\right),
\]
where $\tilde{I}_{\mu}$ indicates the integrated 4pt correlator after
subtracting the divergent contributions.

\subsection{Single Pion Intermediate State}

Our exploratory simulations use an unphysically heavy pion mass of $\sim\unit[420]{MeV}$, and so we worry only
about the single pion divergence; below we discuss the methods
of its removal. A discussion of the $\pi\pi$ and $\pi\pi\pi$ divergences can be found in Ref.~\cite{Christ:2015aha}.

The first possibility is to reconstruct the analytical form of the
divergence from Eq. (\ref{eq:int_spec_rep-1}). The divergent contribution
is therefore
\[
D_{\pi}\left(T_{a},\mathbf{k},\mathbf{p}\right)=\dfrac{1}{2E_{\pi}\left(\mathbf{k}\right)}\dfrac{\mathcal{M}_{J,\pi}^{\mu}\left(\mathbf{q}\right)\mathcal{M}_{H}\left(\mathbf{k}\right)}{E_{K}\left(\mathbf{k}\right)-E_{\pi}\left(\mathbf{k}\right)}e^{\left(E_{K}\left(\mathbf{k}\right)-E_{\pi}\left(\mathbf{k}\right)\right)T_{A}},
\]
where $\mathcal{M}_{\mu}^{J,P}\left(\mathbf{q}\right)=\left\langle P\left(\mathbf{p}\right)|J_{\mu}|P\left(\mathbf{k}\right)\right\rangle $
and $\mathcal{M}_{H}\left(\mathbf{k}\right)=\left\langle \pi\left(\mathbf{k}\right)|H_{W}|K\left(\mathbf{k}\right)\right\rangle $.
The necessary matrix elements and energies can easily be recovered
from fits to 2pt and 3pt correlators. In our exploratory studies the
single pion state is the only divergent state, and so we could alternatively
determine the contribution by fitting the 4pt function directly. We
remark that for the 4pt case we do not fit the exponent of the $T_{a}$
dependence, which we can determine much more reliably from fits to
2pt correlators. We will refer to this divergence subtraction method as "method 1".

A second method ("method 2") of removing the single pion state is to employ a shift
of the weak Hamiltonian by an unphysical scalar density, $\bar{s}d$~\cite{Bai:2014cva}.
We choose a constant $c_{s}$ such that
\[\label{eq:sd_shift}
\left\langle \pi|H_{W}^{\prime}|K\right\rangle =\left\langle \pi|H_{W}|K\right\rangle -c_{s}\left\langle \pi|\bar{s}d|K\right\rangle =0.
\]
By replacing $H_{W}$ by $H_{W}^{\prime}$ in Eq. (\ref{eq:int_spec_rep-1})
it is straightforward to see that the divergent single pion contribution
vanishes. We can show that this shift leaves the total amplitude invariant using the chiral Ward identity
\[
i\left(m_{s}-m_{d}\right)\bar{s}d=\partial_{\mu}V_{\bar{s}d}^{\mu}.
\]


\section{Numerical Results\label{sec:Numerical-Results}}

Our integration of the 4pt correlator requires us to simulate with
a long enough time extent between the kaon source at $t_{K}$, the
current insertion at $t_{J}$ and the pion sink at $t_{\pi}$. This
is necessary to allow the integral to converge to extract the desired
matrix element. To increase the time extent over which we
can integrate the rare kaon 4pt correlator find it is useful to 
consider two 4pt correlators with different values
of $t_{J}$, requiring the computation of additional sequential
propagators. By using a current insertion closer to the pion sink,
we obtain a longer time extent over which to integrate in the region
$\left[t_{J}-T_{A},t_{J}\right]$. Likewise by inserting the current
closer to the kaon source, we obtain a longer time extent to integrate
over $\left[t_{J},t_{J}+T_{B}\right]$. Hence we can obtain the two halves of the integral from these separate correlators and
combine them appropriately to obtain the final result.

We ran a set of simulations with a kaon at $t_{K}=0$ with momentum
$\mathbf{k}$ decaying into a pion at $t_{\pi}=34$ with momentum
$\mathbf{p}$, with two separate current insertions as described above.
We make use of Coulomb gauge-fixed wall sources for the pion and kaon to provide good overlap with
the ground state. For the current insertion we make use of sequential propagators. The
loops for the $S$ and $E$ (and disconnected) diagrams can be calculated using all-to-all
propagators. However for our exploratory calculations we omit these diagrams to decrease the 
simulation time in order to develop our analysis techniques.
Three different kinematics have been studied so far, $\mathbf{k}\rightarrow\mathbf{p}=\left(0,0,0\right)\rightarrow\left(1,0,0\right)$,
$(0,0,0)\rightarrow(1,1,0)$ and $\left(1,0,0\right)\rightarrow\left(0,0,0\right)$ (momentum given as multiples of $2\pi/L$).
All of our results were found using a sample of 256 configurations
of a $24^{3}\times64$ lattice with an inverse lattice spacing of
$1/a=\unit[1.73]{GeV}$, employing domain wall fermions with Iwasaki gauge action, a pion mass of $\sim\unit[420]{MeV}$ and a kaon
mass of $\sim\unit[600]{MeV}$~\cite{Aoki:2010dy}. For the renormalisation of $H_{W}$ we may use the results 
of Ref.~\cite{Christ:2012se}; where a lattice of the same spacing, but smaller volume was used to
perform the necessary NPR. We expect the results to hold for our lattice too as the renormalisation procedure depends
upon the UV behaviour of the theory and thus is insensitive to finite volume effects.

\begin{figure}
\begin{centering}
\begin{tabular}{cc}
\includegraphics[width=7cm]{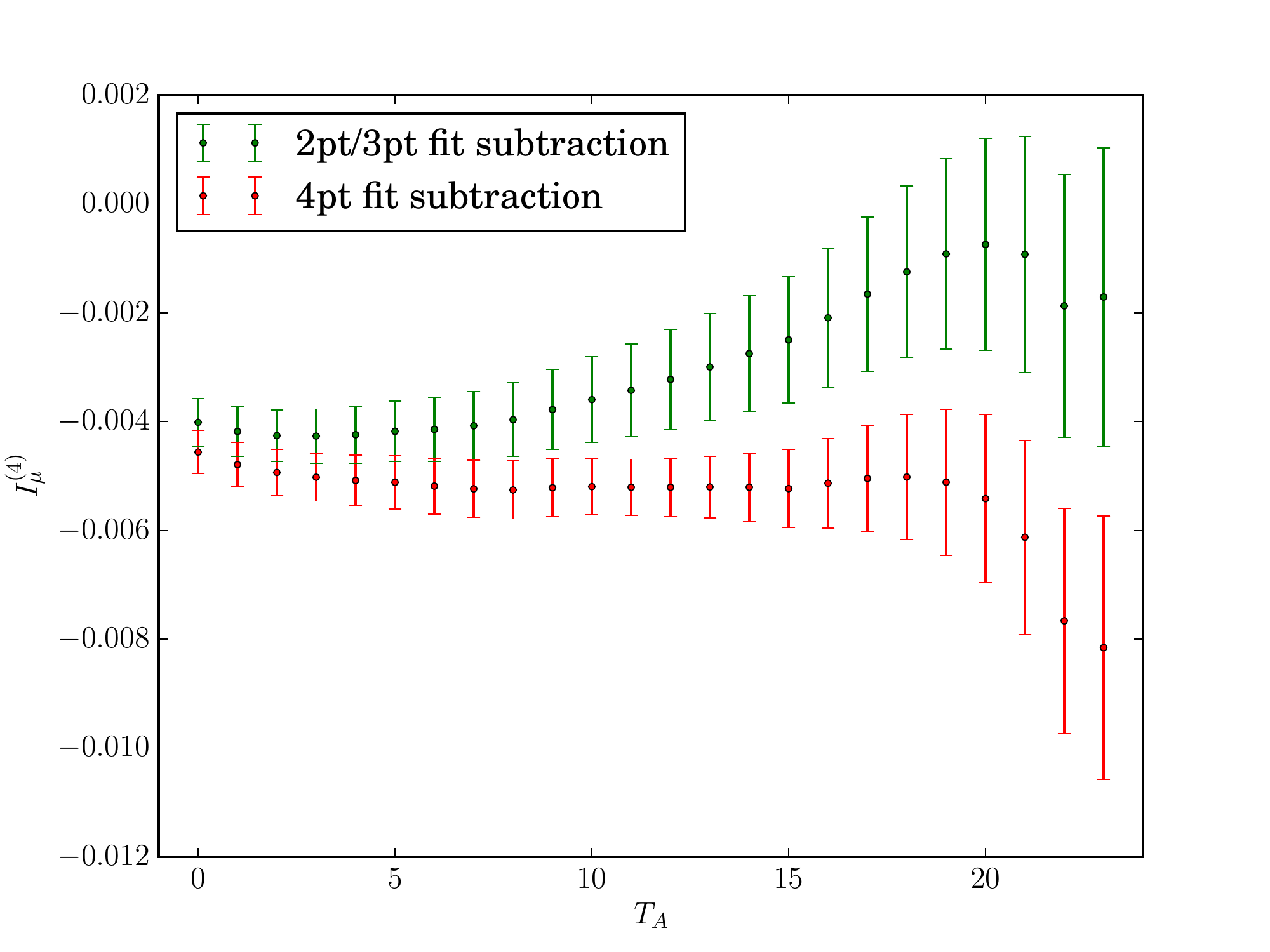} & \includegraphics[width=7cm]{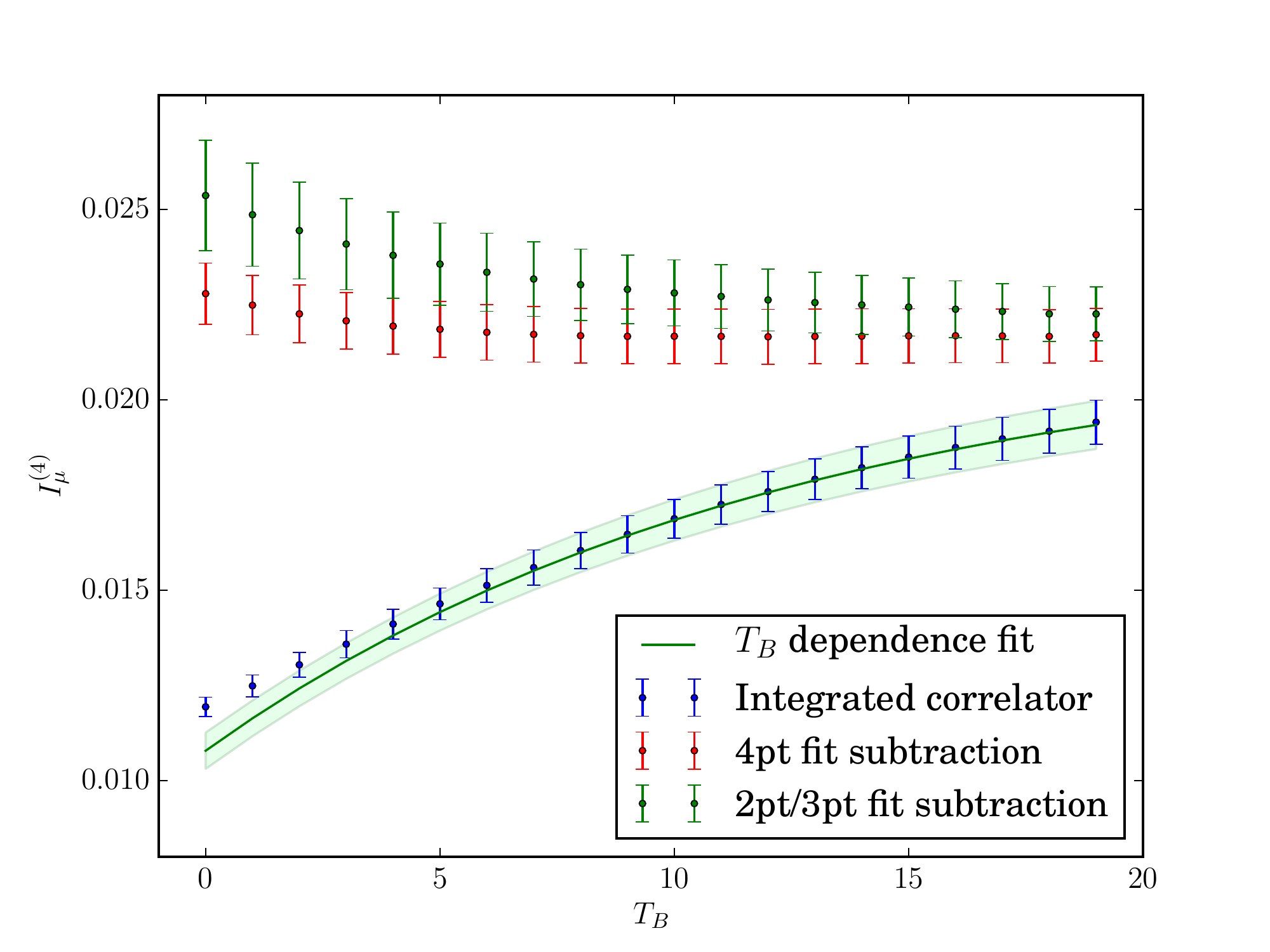}\tabularnewline
$(a)$ & $(b)$\tabularnewline
\end{tabular}
\par\end{centering}

\protect\caption{\label{fig:Method_1_extrap_t_pi=00003D34}The dependence
of the integrated 4pt correlator on the limits $(a)$ $T_{A}$ (with
$T_{B}=19$ fixed) and $(b)$ $T_{B}$ (with $T_{A}=4$ fixed). In $(a)$ the divergence has been removed using method 1.}
\end{figure}

\begin{figure}
\begin{centering}
\begin{tabular}{cc}
\includegraphics[width=7cm]{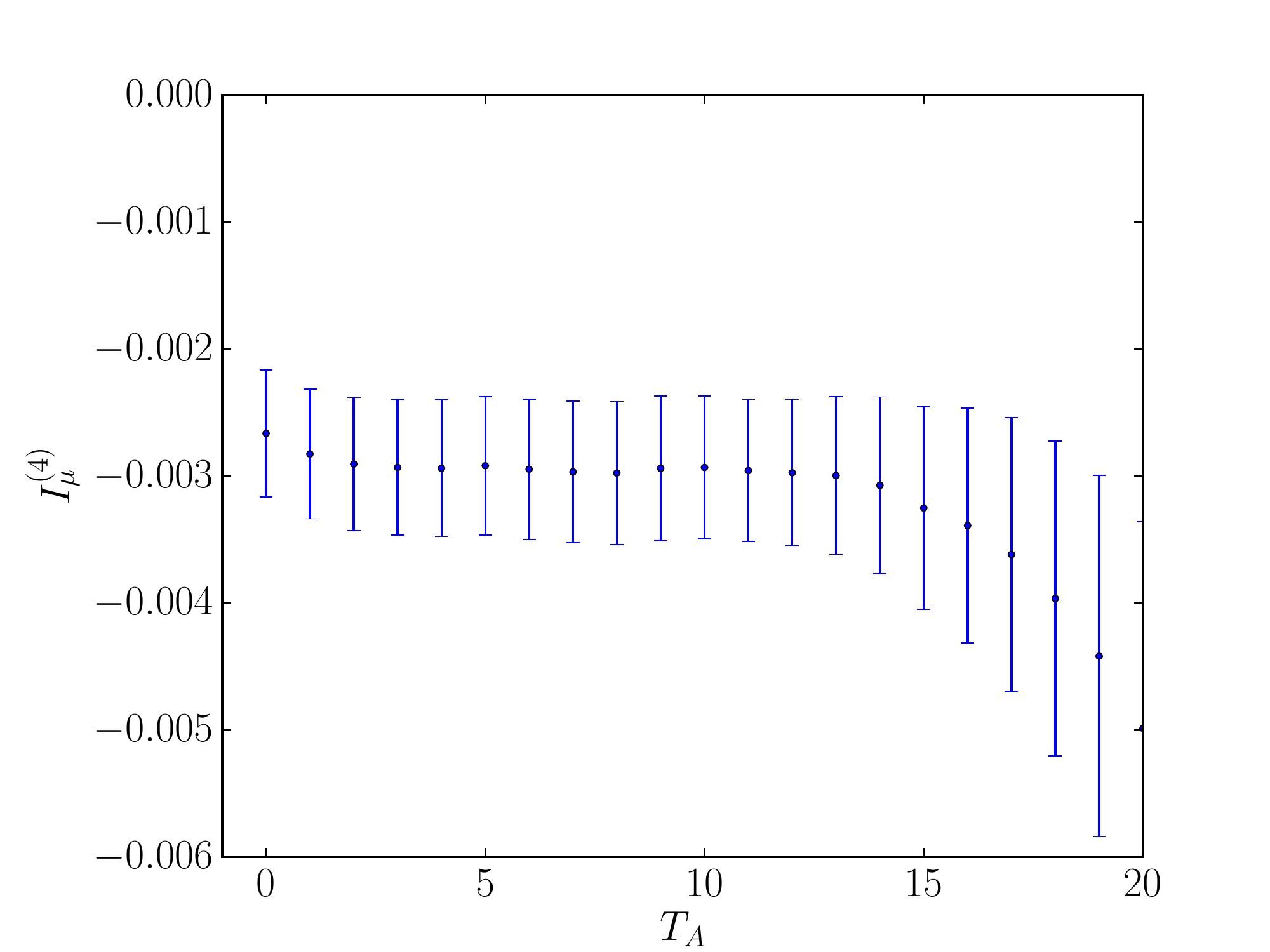} & \includegraphics[width=7cm]{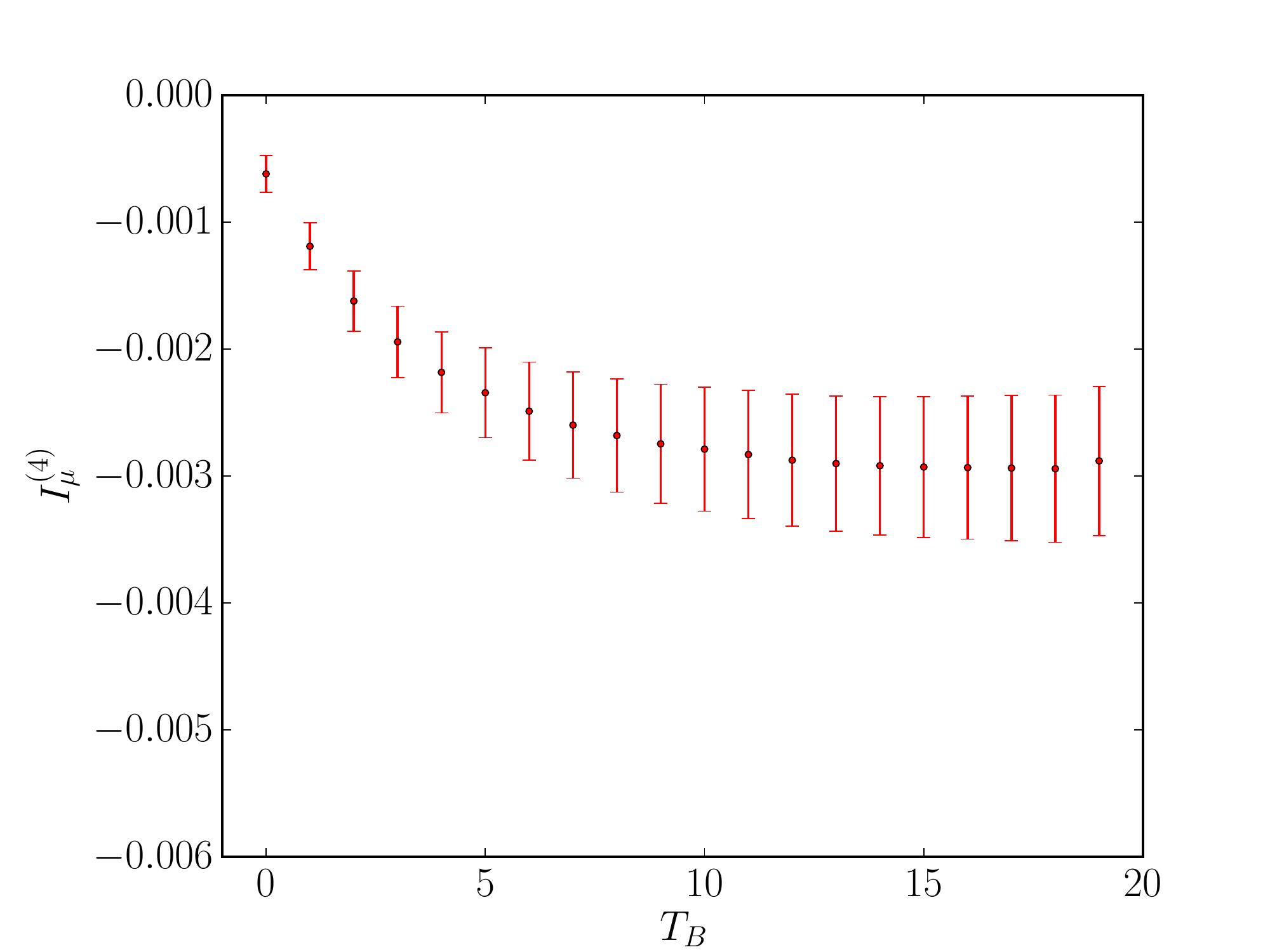}\tabularnewline
$(a)$ & $(b)$\tabularnewline
\end{tabular}
\par\end{centering}

\protect\caption{\label{fig:Method_2_extrap_t_pi=00003D34}The dependence
of the integrated 4pt correlator on the limits $(a)$ $T_{A}$ (with
$T_{B}=14$ fixed) and $(b)$ $T_{B}$ (with $T_{A}=7$ fixed), having
removed the single pion divergence using method 2.}
\end{figure}

We display the dependence of the integrated 4pt correlator on the limits $T_A$ and $T_B$ in Fig.~\ref{fig:Method_1_extrap_t_pi=00003D34}, removing the single pion divergence using method 1. The fit of the 4pt function appears to remove the divergence more cleanly that by using the reconstruction from 2pt/3pt fits. More investigation is necessary to understand the source of this discrepancy.
We also observe that the $T_{B}$ dependence does not converge in the available time extent. which can be attributed to the fact that here the kaon-pion mass difference is rather small, hence a small exponent for the decay. In practice it is thus necessary to remove this decay in a manner similar to the divergence.

We display the dependence of the integral of the 4pt function for the ${K(0,0,0)\rightarrow\pi(1,0,0)}$ kinematic in Fig. \ref{fig:Method_2_extrap_t_pi=00003D34}, where the divergence has been removed using method 2. We note the integral converges to a plateau within statistical errors for both sides of the integral. The parameter $c_s$ has a relatively weak momentum dependence, and so the $\bar{s}d$ shift reduces the amplitude of the exponentially decaying contribution of the single pion state.

\begin{table}
\begin{centering}
\begin{tabular}{|c|c|c|c|}
\hline 
$\mathbf{k}\rightarrow\mathbf{p}$ & Method 1 (4pt fit) & Method 2 ($\bar{s}d$ shift)\tabularnewline
\hline 
$\left(0,0,0\right)\rightarrow\left(1,0,0\right)$ & $-0.0029(6)$ & $-0.0030(6)$\tabularnewline
\hline 
$\left(0,0,0\right)\rightarrow\left(1,1,0\right)$ & $-0.0049(23)$ & $-0.0024(27)$\tabularnewline
\hline 
$\left(1,0,0\right)\rightarrow\left(0,0,0\right)$ & $-0.0018(11)$ & $-0.0001(7)$ \tabularnewline
\hline 
\end{tabular}
\par\end{centering}

\protect\caption{\label{tab:Summary}The $\mu=0$ component of the matrix element (in lattice units) found with each analysis method.}
\end{table}

In Table \ref{tab:Summary} we present a summary of the results for the two methods of extracting the matrix element from the integrated 4pt correlator. The two methods generally agree within statistical errors. We remark that we can fit the 4pt function with an $\bar{s}d$ insertion in a similar manner to the original in order to verify that after removing the divergence the integral of the correlator is within error of zero. For the $K(1,0,0)\rightarrow\pi(0,0,0)$ case we identify that the contribution of the integrated $\bar{s}d$ correlator is more than one sigma different from zero. More investigation is required to identify the source of any potential systematic effects in our analysis procedure.

\begin{figure}
\begin{centering}
\includegraphics[width=7cm]{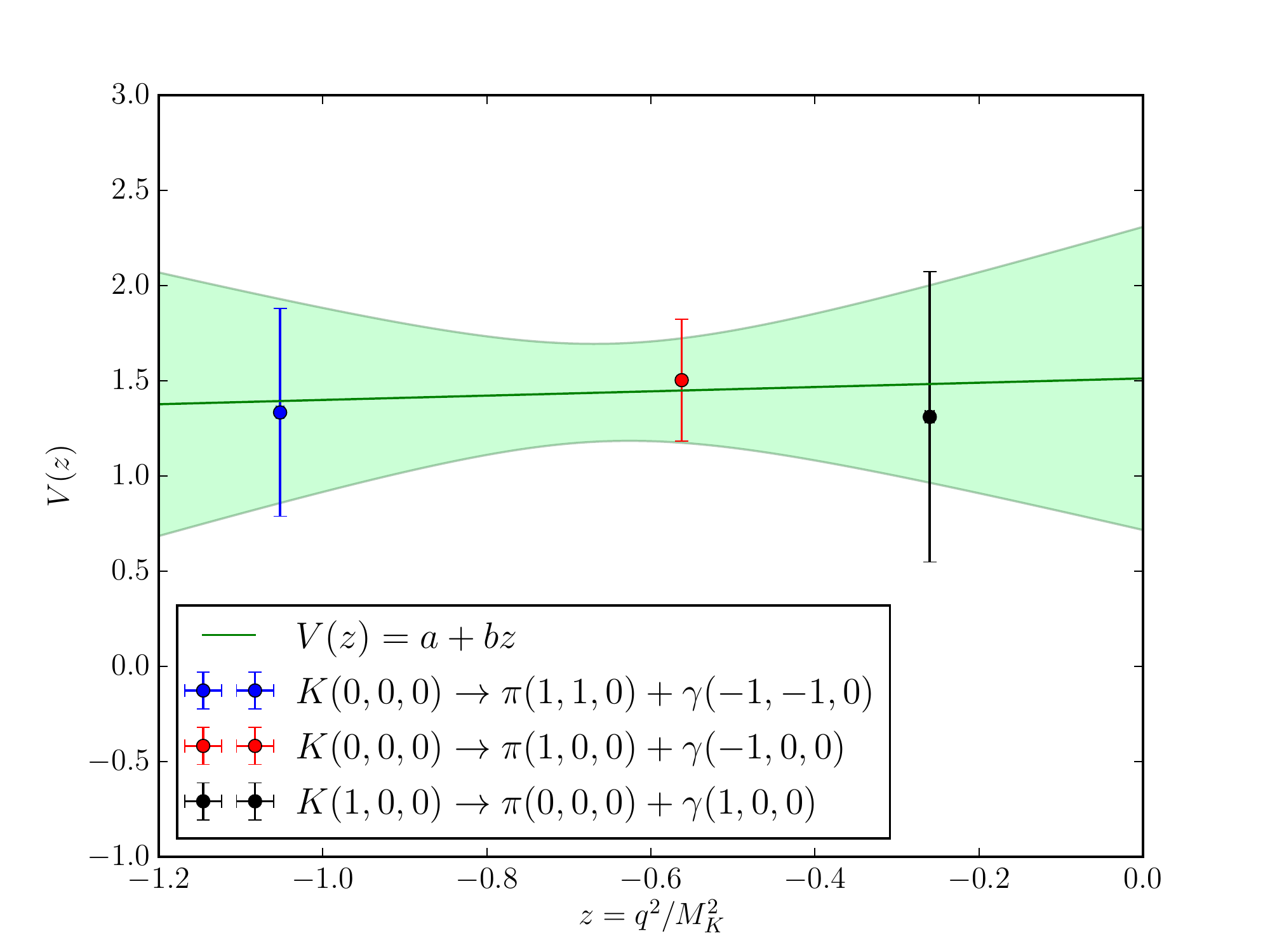}
\par\end{centering}

\protect\caption{\label{fig:form_factor}The dependence of the form factor on $q^2$ from our lattice simulations.}

\end{figure}

Finally we are able to use our results to determine the form factor for each
kinematic. These results are shown in Fig. \ref{fig:form_factor}, where 
we have used the matrix element obtained by fitting the 4pt function
directly. The fit ansatz is motivated by the ChPT prediction; we omit the $\pi\pi\rightarrow\gamma^{*}$ contribution here
as it is negligible. At our current level of statistics the form factor does not appear to display any discernible dependence on $q^2$. We obtain the fit parameters $a=1.5(8)$ and $b=0.1(1.1)$. For contrast the experimentally determined parameters are $a=-0.578(16)$, $b=-0.779(66)$ for $K^+\rightarrow\pi^+e^+e^-$ and $a=-0.575(39)$, $b=-0.813(145)$ for $K^+\rightarrow\pi^+\mu^+\mu^-$~\cite{Cirigliano:2011ny}. The comparison of central values is meaningless given our unphysical parameters, however we note that our errors are an order of magnitude greater than experiment. A significant gain in statistics would be needed before our errors become competitive.

\section{Conclusions\label{sec:Conclusions}}

Through our exploratory numerical simulations we have evaluated the different analysis techniques for extracting the long distance contributions to the decay $K^+\rightarrow\pi^+\ell^+\ell^-$ using lattice QCD. At present remains to extend our simulations to include the full set of diagrams on the current lattice. We subsequently aim to study the decay with more physical kaon/pion masses where it will be important to also consider $\pi\pi$ or even $\pi\pi\pi$ divergent contributions.

\section*{Acknowledgements}

A.L is supported by an EPSRC Doctoral Training Centre grant (EP/G03690X/1). N.H.C and X.F are supported by US DOE grant \#DE-SC0011941. A.P and C.T.S are supported by UK STFC Grant ST/L000296/1. A.J acknowledges the European Research Council under the European Community's Seventh Framework Programme (FP7/2007-2013) ERC grant agreement No. 279757.

\bibliography{proceedings}

\providecommand{\href}[2]{#2}\begingroup\raggedright\begin{thebibliography}{1}

\bibitem{Cirigliano:2011ny}
V.~Cirigliano, G.~Ecker, H.~Neufeld, A.~Pich, and J.~Portoles, {\it {Kaon
  Decays in the Standard Model}},  {\em Rev. Mod. Phys.} {\bf 84} (2012) 399,
  [\href{http://arxiv.org/abs/1107.6001}{{\tt arXiv:1107.6001}}].

\bibitem{Isidori:2005tv}
G.~Isidori, G.~Martinelli, and P.~Turchetti, {\it {Rare kaon decays on the
  lattice}},  {\em Phys. Lett.} {\bf B633} (2006) 75--83,
  [\href{http://arxiv.org/abs/hep-lat/0506026}{{\tt hep-lat/0506026}}].

\bibitem{Christ:2015aha}
{\bf RBC, UKQCD} Collaboration, N.~H. Christ, X.~Feng, A.~Portelli, and C.~T.
  Sachrajda, {\it {Prospects for a lattice computation of rare kaon decay
  amplitudes: $K\to\pi\ell^+\ell^-$ decays}},  {\em Phys. Rev.} {\bf D92}
  (2015) 094512, [\href{http://arxiv.org/abs/1507.03094}{{\tt
  arXiv:1507.03094}}].

\bibitem{Ecker:1987qi}
G.~Ecker, A.~Pich, and E.~de~Rafael, {\it {K ---$>$ pi Lepton+ Lepton- Decays
  in the Effective Chiral Lagrangian of the Standard Model}},  {\em Nucl.
  Phys.} {\bf B291} (1987) 692.

\bibitem{D'Ambrosio:1998yj}
G.~D'Ambrosio, G.~Ecker, G.~Isidori, and J.~Portoles, {\it {The Decays K ---$>$
  pi l+ l- beyond leading order in the chiral expansion}},  {\em JHEP} {\bf
  9808} (1998) 004, [\href{http://arxiv.org/abs/hep-ph/9808289}{{\tt
  hep-ph/9808289}}].

\bibitem{Buchalla:1995vs}
G.~Buchalla, A.~J. Buras, and M.~E. Lautenbacher, {\it {Weak decays beyond
  leading logarithms}},  {\em Rev. Mod. Phys.} {\bf 68} (1996) 1125--1144,
  [\href{http://arxiv.org/abs/hep-ph/9512380}{{\tt hep-ph/9512380}}].

\bibitem{Bai:2014cva}
Z.~Bai, N.~Christ, T.~Izubuchi, C.~Sachrajda, A.~Soni, et~al., {\it {$K_L-K_S$
  mass difference from lattice QCD}},  {\em Phys. Rev. Lett.} {\bf 113} (2014)
  112003, [\href{http://arxiv.org/abs/1406.0916}{{\tt arXiv:1406.0916}}].

\bibitem{Aoki:2010dy}
{\bf RBC, UKQCD} Collaboration, Y.~Aoki et~al., {\it {Continuum Limit Physics
  from 2+1 Flavor Domain Wall QCD}},  {\em Phys. Rev.} {\bf D83} (2011) 074508,
  [\href{http://arxiv.org/abs/1011.0892}{{\tt arXiv:1011.0892}}].

\bibitem{Christ:2012se}
{\bf RBC, UKQCD} Collaboration, N.~Christ, T.~Izubuchi, C.~Sachrajda, A.~Soni,
  and J.~Yu, {\it {Long distance contribution to the KL-KS mass difference}},
  {\em Phys. Rev.} {\bf D88} (2013) 014508,
  [\href{http://arxiv.org/abs/1212.5931}{{\tt arXiv:1212.5931}}].

\end{thebibliography}\endgroup
\bibliographystyle{JHEP}

\end{document}